# Fabrication of nanotubules of thermoelectric $\gamma$-$Na_{0.7}CoO_2$ using porous aluminum oxide membrane as supporting template

*Chia-Jyi Liu,* [*] *Shu-Yo Chen, and Long-Jiann Shih*

Department of Physics, National Changhua University of Education, Changhua 50007, Taiwan, R. O. C.

AUTHOR EMAIL ADDRESS: ncuecj@yahoo.com

## ABSTRACT

*We report the successful synthesis of nanotubules of thermoelectric materials $\gamma$-$Na_xCoO_2$ using two different sol-gel routes aided by porous anodized aluminium oxide (AAO) membrane as supporting templates. The $\gamma$-$Na_xCoO_2$ nanotubule using urea-based route can be achieved at 650ºC at a heating rate of 1ºC/min and held for 4h. The $\gamma$-$Na_xCoO_2$ nanotubule using citric acid-based route can be achieved at 500ºC using a rapid-heat-up procedure and held for 30 min. The products were investigated using various techniques including XRD, SEM and TEM. Electron diffraction pattern taken along [001] zone axis direction on the nanotubule shows that all the diffraction spots can be indexed using a hexagonal unit cell with a = b = 0.56 nm, which can be considered as a superstructure with cell doubling within the* ab *plane.*

KEYWORDS: Nanomaterial, Thermoelectric oxide, AAO template aided sol-gel synthesis.

Thermoelectric materials hold great promise for clean energy generation by transforming heat into electricity due to the Seebeck effect. The γ-phase of $Na_xCoO_2$ is a potential candidate for thermoelectric applications due to its high electrical conductivity, large thermopower and low thermal conductivity. [1,2] Figure 1 shows the structure of γ-$Na_{0.7}CoO_2$, which has a hexagonal lattice with the space group of $P6_3/mmc$ (#194), consisting of two $CoO_2$ layers with the Na ions located in two different crystallographic sites between them. The c-axis is ca. 1.09 nm and therefore has an interlayer spacing of ca. 0.545 nm between the two $CoO_2$ layers. In addition, γ-$Na_{0.7}CoO_2$ can be topotactically transformed to superconducting oxyhydrates $Na_x(H_2O)_yCoO_2$ with $T_c$ = 4.6 K. [3,4]

Synthesis of one-dimensional nanomaterials with novel properties has attracted significant attention in the recent years. Nanoporous membranes can be used as templates to prepare nanocylindrical materials such as solid nanofibrils or hollow nanotubules. [5–7] A significant increase of the thermoelectric figure of merit could be achieved for one-dimensional conductors or quantum wires through enhancing the thermopower by increasing $\partial g(E)/\partial E$ near the Fermi energy, where g(E) is the density of states, and lowering the lattice thermal conductivity by increasing the boundary phonon scattering. [8] This promise would not be fulfilled without making the required one-dimensional structure. Herein, we demonstrate fabrication of nanotubes of thermoelectric oxides γ-$Na_xCoO_2$ using two different sol-gel routes with the assistance of porous anodized aluminum oxide membrane as supporting templates.

The γ-$Na_xCoO_2$ nanotubules were synthesized by the urea-based sol-gel route [9] with the assistance of supporting templates using anodized aluminum oxide (AAO) membranes having a pore size ranging from 40-350 nm. The sol used in the fabrication of nanotubules was prepared by dissolving metal nitrates in a molar ratio of $NaNO_3$ : $Co(NO_3)_3·6H_2O$ = 1 : 1 (0.05 mol each) in 50 ml of deionized water. Urea (0.133 mol) was then added in a molar ratio of Co : urea = 3 : 8 and heated to 60- 80ºC on a hot plate until the sol was formed. The AAO membranes were dipped into the sol for 10 min. The AAO membrane was taken out and filled with the hot sol gradually using vacuum immersion techniques for a period of 2 h. The filled AAO membrane was then dried on a hot plate at 65ºC for 30 min, followed by cleaning its surface using anhydrous alcohol to remove residual gel. The resulting filled AAO membrane was then heated to 600ºC at a heating rate of 1ºC/min and held for 10 min, followed by heating to 650ºC at a heating rate of 1ºC/min and held for 4 h. The reaction temperature was purposely to be kept under 700ºC since AAO membrane after being heated above 700ºC would persistently remain in the final product even if treated with alkaline NaOH solution. The desired phase can be obtained by heating the filled membranes at as low as 650°C. Note that polycrystalline $\gamma$-$Na_{0.7}CoO_2$ powders were normally synthesized in a preheated furnace at 800°C.

The $\gamma$-$Na_xCoO_2$ nanotubules were characterized using a Shimadzu XRD-6000 powder x-ray diffractometer equipped with Fe K$\alpha$ radiation (XRD), a JEOL JEM-2010 transmission electron microscope (TEM) equipped with OXFORD INCAx-sight energy dispersive x-ray (EDX) spectrometer detector and INCAx-stream pulse processor, and a JEOL JSM-6700F field-emission scanning electron microscope (FE-SEM) equipped with Oxford INCA Energy 350 energy dispersive X-ray spectrometer.

Figures 2a and 2b show the SEM images of $\gamma$-$Na_xCoO_2$ nanotubules synthesized by a urea-based sol-gel route using the AAO membrane as supporting template. Since the pore size of the AAO membrane is not uniform, the as-prepared nanotubules from an AAO membrane with a pore size of 230-370 nm has a diameter size ranging from 200 nm to 340 nm (see Figure 2a). The EDX spectra taken over about 20 nanotubules reveal that the atomic ratio of Na/Co for the nanotubules is 0.72, which corresponds to the chemical composition of $Na_{0.72}CoO_2$. The nanotubules (see Figure 2b) ranging from 35 to 97 nm in diameter are obtained using an AAO membrane with a pore size ranging from 40 to 100 nm. The SEM images demonstrate that most nanotubules have a uniform diameter along their length. EDX analysis is also carried out on a single nanotubule with a diameter of about 40 nm (see Figure 2c) and reveals that the atomic ratio of Na/Co for the nanotubule is 0.7, which corresponds to the chemical composition of $Na_{0.7}CoO_2$. Note that due to the nature of volatility of Na in the course of heating, the atomic ratio of Na/Co is less than the starting ratio of Na/Co = 1 in the precursor solution. Figure 2d shows the selected area electron diffraction (SAED) pattern of the γ-$Na_xCoO_2$ nanotubule taken along the [001] zone axis direction at room temperature. All the diffraction spots can be indexed using a hexagonal unit cell with $a = b = 0.56$ nm. In-situ heating transmission electron microscopy observations of $Na_xCoO_2$ indicate that there appears a notable superstructure with ordering between $CoO_2$ layers. [10] In addition, complex superstructures due to the sodium ion ordering are observed in the layered sodium cobalt oxides $Na_xCoO_2$. [11] Therefore, the observed cell parameters with $a = b = 0.56$ nm for the γ-$Na_xCoO_2$ nanotubule could be considered as a superctructure having cell doubling in the *ab* plane.

In order to see if the nanotubules crystallize in the hexagonal γ-$Na_xCoO_2$ phase in a large quantity, we show the XRD pattern (Figure 3) of nanotubules, which are separated from the nanotubules by dissolving in 0.1 M NaOH at 80°C. Since a very small amount of the nanotubules can be collected in a batch using one piece of AAO membrane, which could not allow the XRD pattern to be observed clearly. The XRD pattern of γ-$Na_xCoO_2$ nanotubules was obtained using 20 pieces of AAO membrane in one batch. In the same graph for comparison, we also show the XRD pattern of bulk powder sample (Figure 2b) prepared using a rapid-heat-up procedure. Except two additional peaks of $2\theta = 48.82°$ and $59.22°$, the remaining diffraction peaks of XRD pattern for the nanotubules can be indexed by a hexagonal unit cell using the same space group of $P6_3/mmc$ as bulk crystalline powders, which confirms that the nanotubules crystallize in the same hexagonal lattice as γ-$Na_xCoO_2$.

In addition, an alternative route to synthesize γ-$Na_xCoO_2$ nanotubules can be assisted using citric acid as the chelating agent with the aid of a porous Whatman AAO membrane having a nominal pore size of 100 nm. The preparation procedure of citric acid-based route is slightly different from the urea-based route and is described briefly as follows. The sol used in the fabrication of nanotubules was prepared by dissolving the metal nitrates in a molar ratio of $NaNO_3$ ; $Co(NO_3)_3·6H_2O = 0.7 : 1$ (0.035 mol : 0.05 mol) in 50 ml of deionized water. Citric acid (0.0.085 mol) was then added in the above solution and heated to 60 - 80°C on a hot plate. The AAO membranes were dipped into the sol for 10 min. The filled AAO membrane was then dried at 65°C for 30 min, followed by cleaning its surface using a piece of Kimwipes wiper. The resulting filled AAO membrane was then heated in a preheated furnace at 500ºC for 30 min using a rapid-heat-up temperature profile. [12] The products were then obtained by repeating the above procedure 3 times. Figure 4a shows the SEM image of γ-Na$_x$CoO$_2$ nanotubules with a diameter of about 200-350 nm. The inset of Figure 4a clearly shows a nanotubule with a diameter of 220 nm. It should be noticed that the size of the diameter of the nanotubule is much larger than the nominal pore size of 100 nm of the Whatman AAO membrane according to the SEM observation. Figure 4b shows a TEM image of the nanotubule with an even larger diameter of 370 nm. The EDX analysis on the nanotubule shown in Figure 4b indicates an atomic ratio of Na/Co = 0.51, which corresponds to the chemical composition of Na$_{0.7}$CoO$_2$.

In summary, we have successfully synthesized nanotubules of thermoelectric materials γ-Na$_x$CoO$_2$ using two different sol-gel routes with the aid of porous AAO membranes as supporting template for the first time. Characterization of the transport properties of the γ-Na$_x$CoO$_2$ nanotubules is in progress.

FIGURE CAPTIONS

Figure 1. Crystal structure of hexagonal $\gamma$-$Na_{0.7}CoO_2$.

Figure 2. (a) SEM image of $\gamma$-$Na_xCoO_2$ nanotubules synthesized using a pore size with a range of 230-370 nm AAO membrane as supporting template. (b) SEM image of $\gamma$-$Na_xCoO_2$ nanotubules synthesized using an AAO membrane with a pore size with a range of 40-100 nm as supporting template. (c) TEM image of a $\gamma$-$Na_xCoO_2$ nanotubule. (d) Electron diffraction pattern taken along [001] zone axis direction on the nanotubule shown in (c). All the diffraction spots can be indexed using a hexagonal unit cell with $a = b = 0.56$ nm, which can be considered as a superstructure with cell doubling within the *ab* plane.

Figure 3. XRD patterns of $\gamma$-$Na_xCoO_2$ (a) nanotubules obtained using 20 pieces of AAO membrane as templates; (b) powders prepared using a rapid-heat-up method.

Figure 4. SEM image of $\gamma$-$Na_xCoO_2$ nanotubules synthesized using citric acid-based sol-ge route aided by Whatman's AAO membrane with a nominal pore size of 100 nm. The inset clearly shows a nanotubule with a diameter of 220 nm. (b) TEM image of a nanotubue having a diameter of 370 nm, which has a chemical composition of $Na_{0.51}CoO_2$.

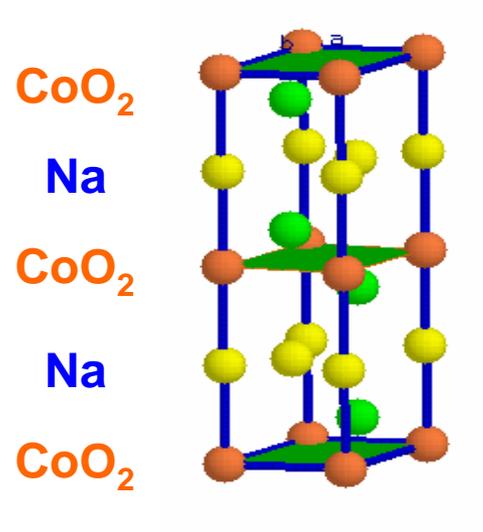

**CoO₂**

**Na**

**CoO₂**

**Na**

**CoO₂**

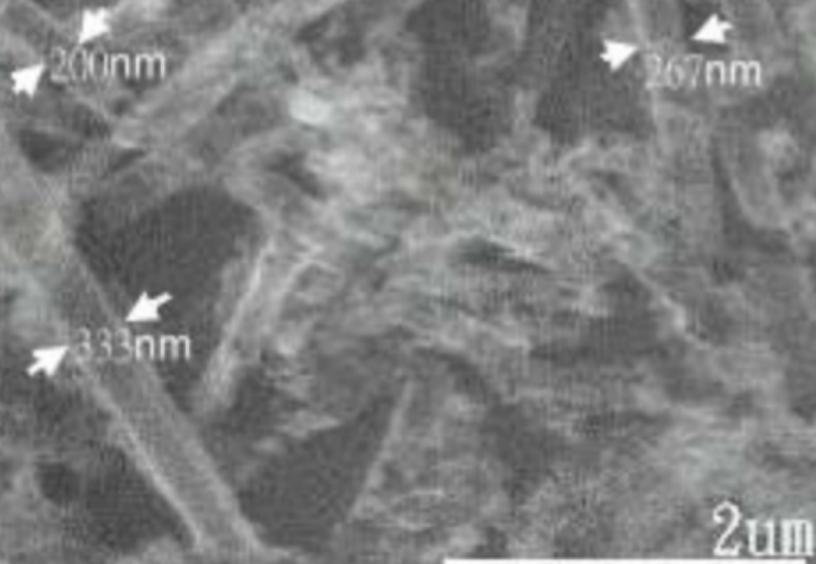

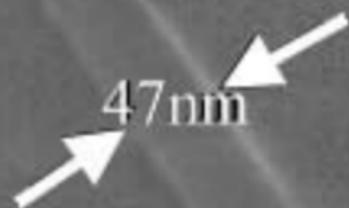
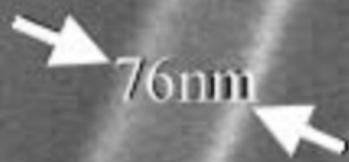
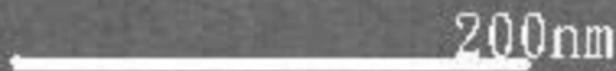

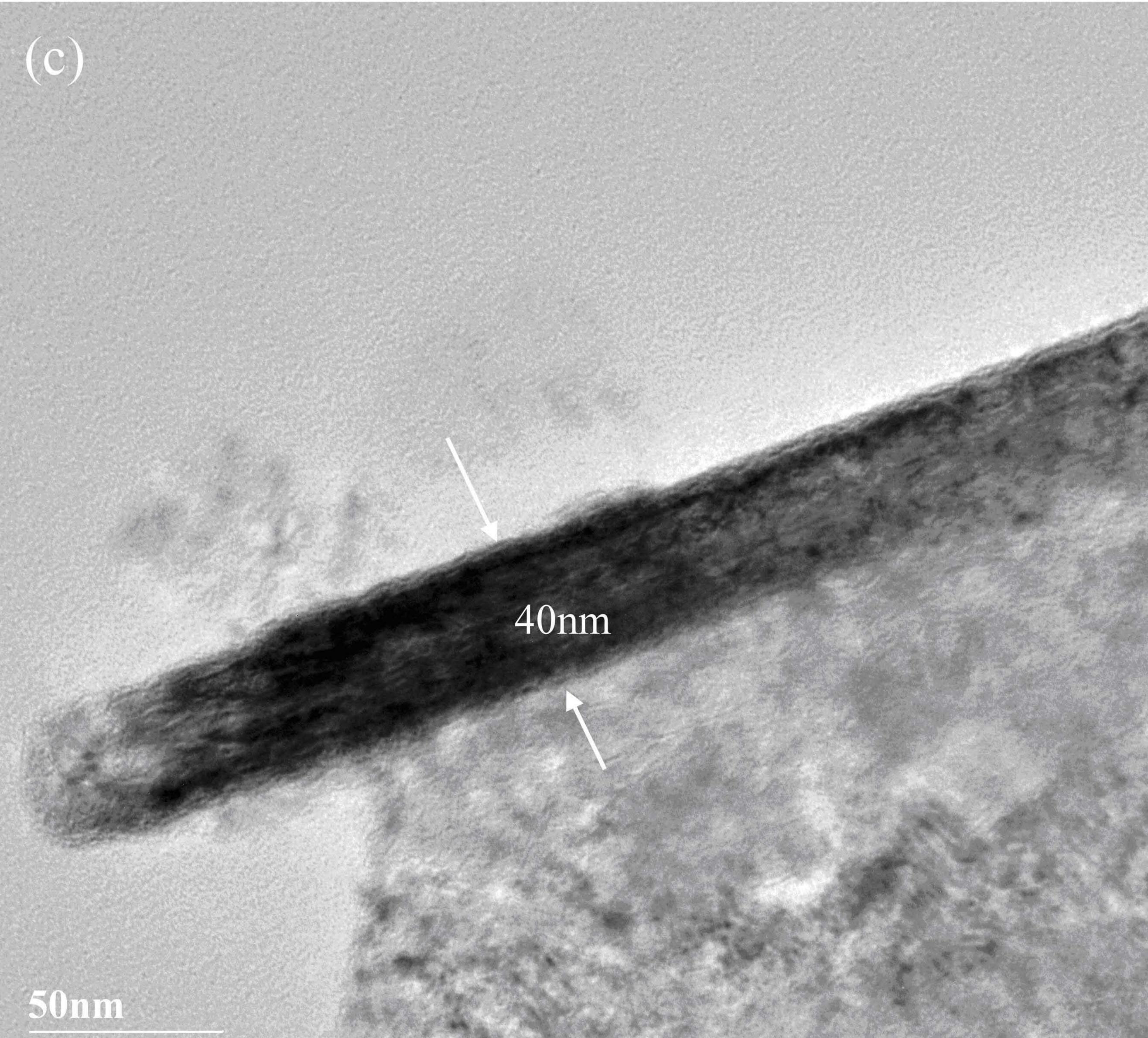

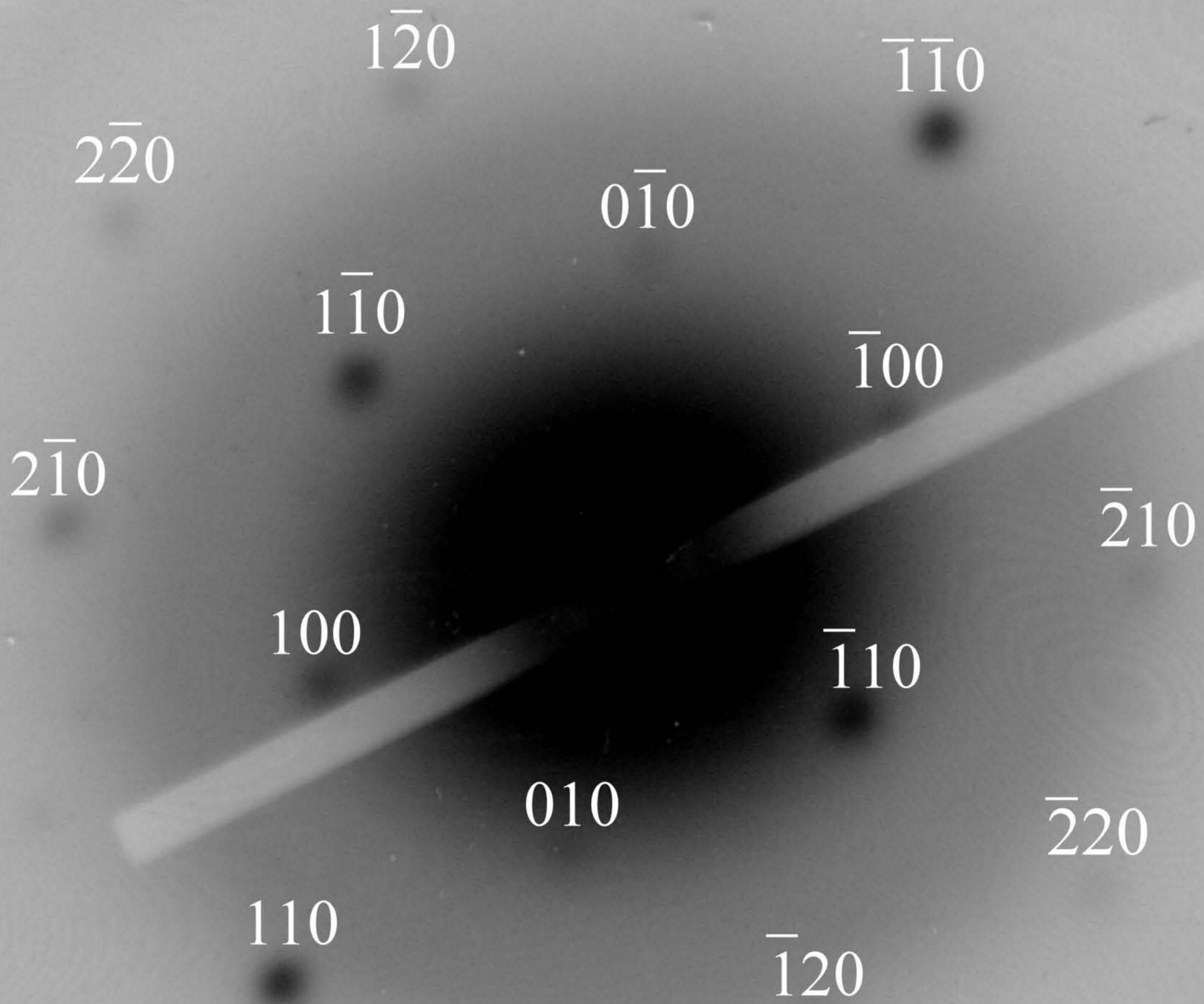

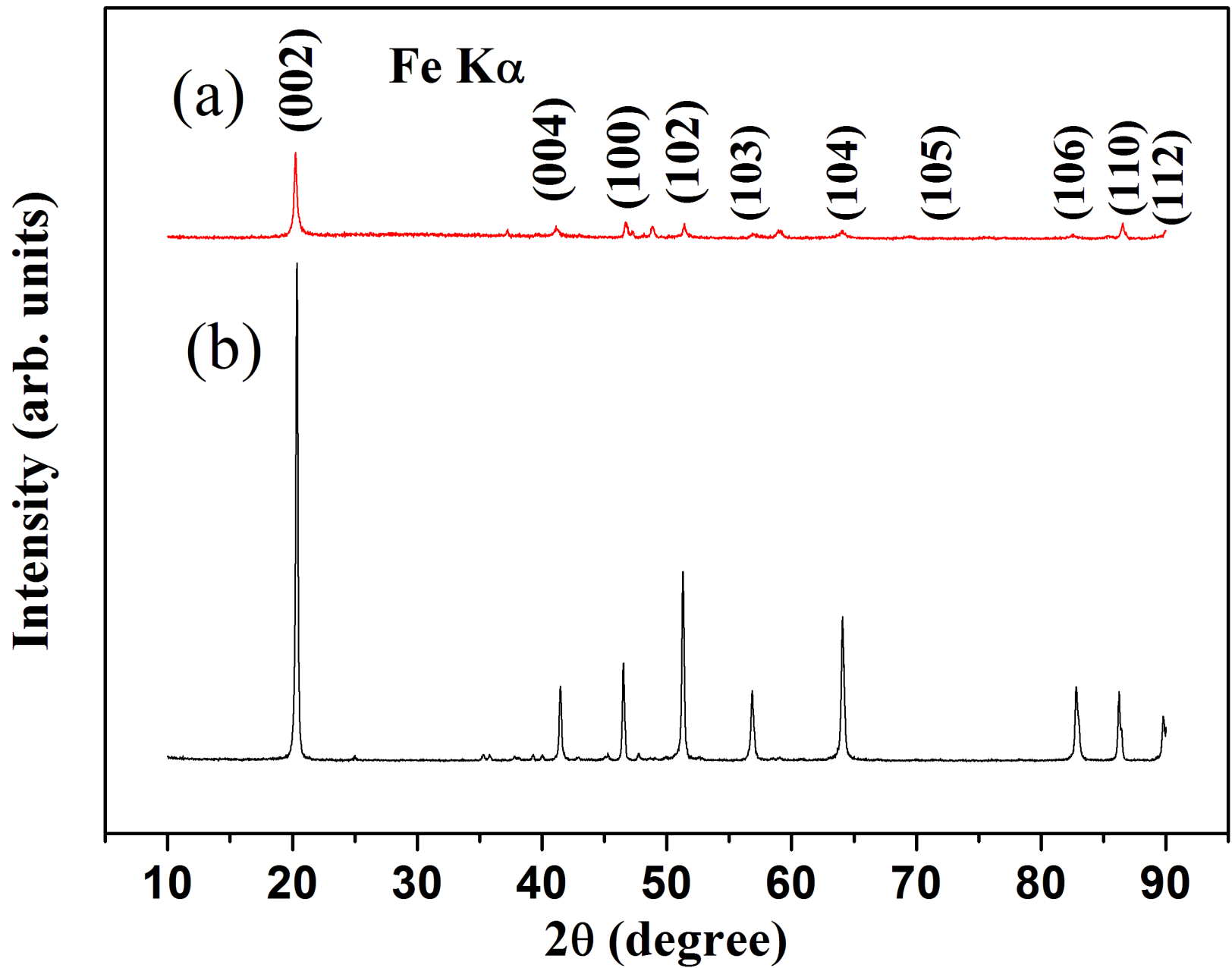

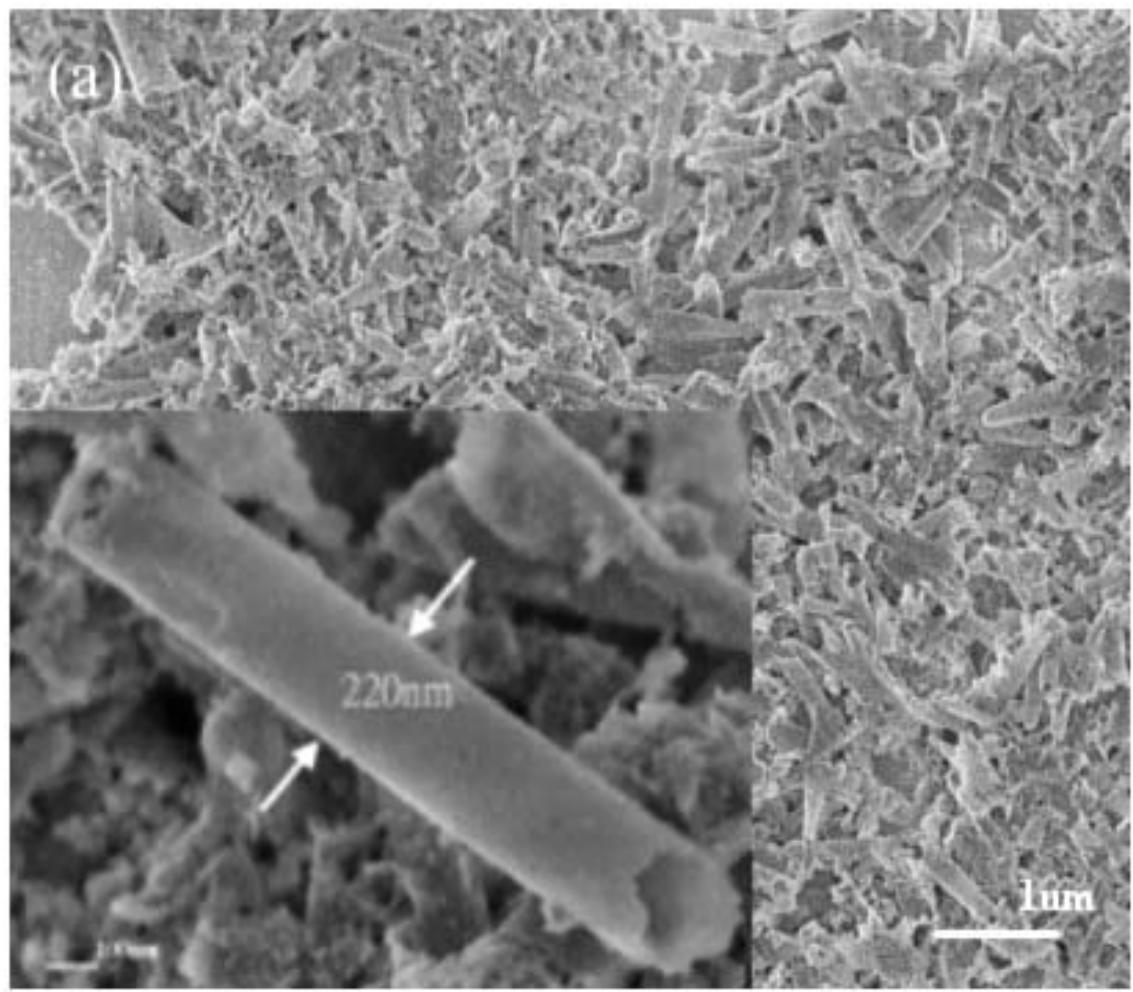

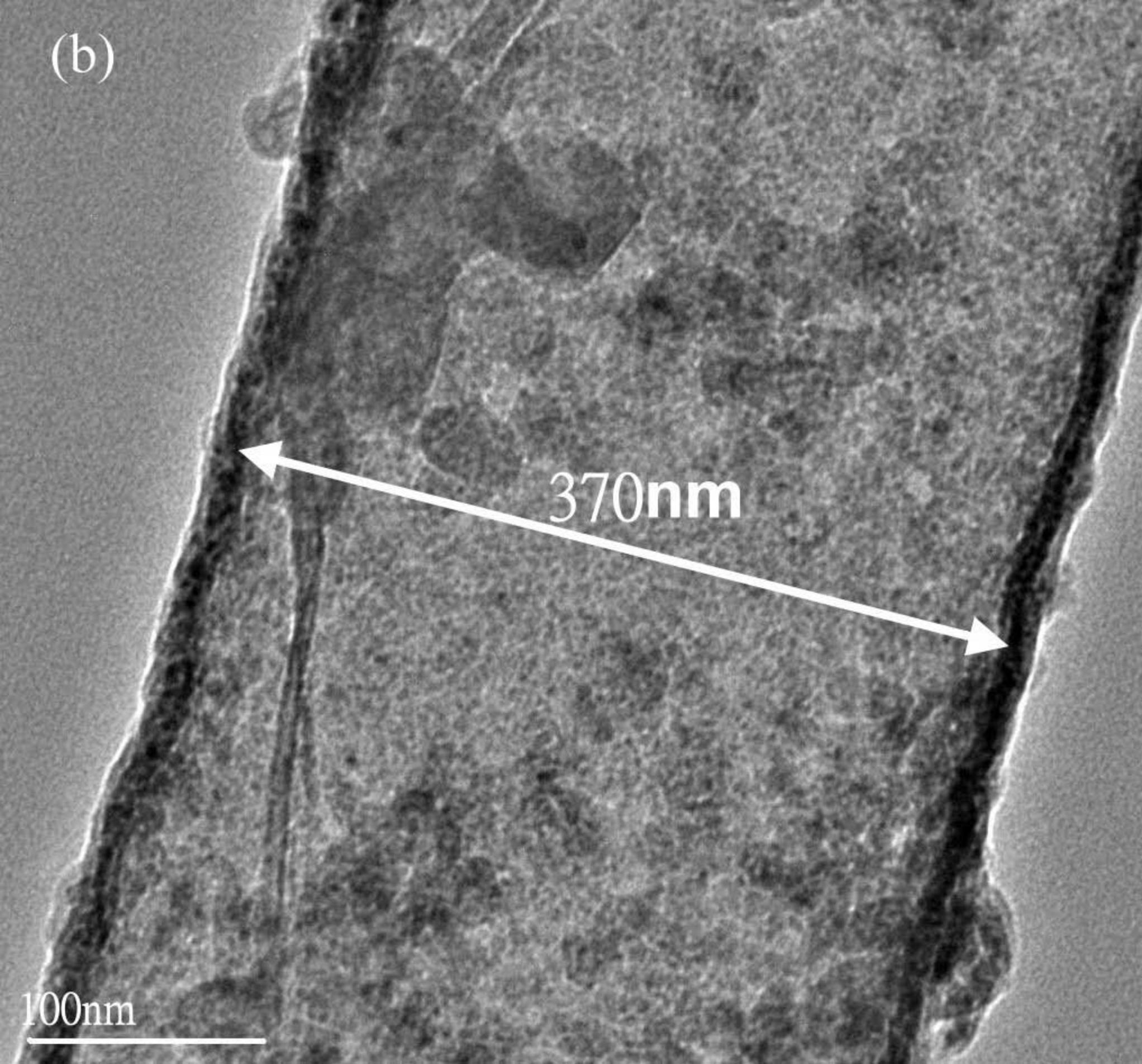